\documentclass[preprint,3p,times,onecolumn]{elsarticle}
\usepackage{amsmath}
\usepackage{epsfig}
\usepackage[titletoc]{appendix}
\usepackage{float}
\usepackage{rotating}
\usepackage{titlesec}
\usepackage{amssymb,graphicx}
\usepackage{amsmath,bm}
\usepackage{multirow}
\newfont{\logo}{logo10}

\usepackage[hang,nooneline]{subfigure}
\newcounter{fig}   
\usepackage{multirow}
\usepackage{hhline}

\newcommand{\bea}{\begin{eqnarray}}
\newcommand{\eea}{\end{eqnarray}}
\newcommand{\bes}{\begin{subequations}}
\newcommand{\ees}{\end{subequations}}

\begin{document}
\begin{frontmatter}
\title{Superposed nonlinear waves in coherently coupled Bose-Einstein condensates}

\author{R. Babu Mareeswaran}
\author{T. Kanna\corref{tk}}\ead{kanna\_phy@bhc.edu.in}
%\author{K. Sakkaravarthi}
\address{Post Graduate and Research Department of Physics, Bishop Heber College, Tiruchirapalli--620 017, Tamil Nadu, India}
\cortext[tk]{Corresponding author}

\date{\today}

\begin{abstract}

We study the dynamics of superposed nonlinear waves in coherently coupled Gross-Pitaevskii (CCGP) equations with constant (autonomous system) and time varying (non-autonomous system) nonlinearity co-efficients. By employing a linear transformation, the autonomous CCGP system is converted into two separate scalar nonlinear Schr\"odinger equations and we show that linear superposition of different nonlinear wave solutions of these scalar equations results into several kinds of nonlinear coherent structures namely, coexisting rogue wave-Ma breather, Akhmediev-Ma breathers, collision and bound states of Ma breathers and solitons. Next, the non-autonomous CCGP system is converted into an autonomous CCGP system with a similarity transformation. We show an interesting possibility of soliton compression and appearance of creeping solitons for kink-like and periodically modulated nonlinearity coefficient.
\end{abstract}

%\pacs{05.45.Yv, 02.30.Ik, 03.75.Lm, 67.85.Hj}
\end{frontmatter}
\section{Introduction}

Nonlinear wave phenomena are of wide physical and mathematical interest as they arise in diverse areas of science such as nonlinear optics, fluid dynamics,
lattice dynamics, plasma physics, bio-physics and Bose-Einstein condensates (BEC) \cite{fluids,ablowitz,wiley,pana}.
In the context of BECs several kinds of nonlinear coherent structures,
such as, solitons \cite{dj}, discrete breathers \cite{dbrea}, rouge waves \cite{bludov}, Faraday waves \cite{engels}, gap solitons \cite{kivshar}
and vortices \cite{vortex} have been observed. Particularly, the multicomponent BECs host rich dynamical behaviors because of the vector order parameters. These systems display interesting effects
such as Josephson effect \cite{joseph}, Efimov effect \cite{efimov}, spin textures \cite{spintext},
spin-orbit coupling \cite{spinorbit}, just to cite a few of them. The fundamental multicomponent BEC is the two component BEC. Such two-component BEC has been realized by using a mixture of atoms with two hyperfine states of
$^{87}Rb$ \cite{rabi11}. The theoretical studies on spin dipole oscillations \cite{rabi1}, vortex pairs \cite{rabi2}, phase transition \cite{abad}, localized waves \cite{pair} and bosons in optical lattices \cite{mole} emphasize the need for investigation of BECs with coherent (phase-dependent) coupling.\\

Observing solitons in multicomponent condensates is one of the current research topics. The formation and dynamical properties of solitons in BECs are determined by the nature of their two-body atomic interactions, i.e., the sign of the $s$-wave scattering length which may be positive (negative) for repulsive (attractive) interatomic interactions. The $s$-wave scattering length and hence the nonlinearity co-efficient can be controlled by means of Feshbach resonance \cite{fesh}. If we use a time-dependent magnetic field, the strength of  the nonlinear interaction can be tuned by time-dependent Feshbach resonance. In real experiments, various forms of time-dependence of nonlinearity have been explored \cite{greiner}.
 In theoretical studies, many authors have thoroughly studied the dynamics of BECs for several forms of time (or) spatial dependent nonlinearity co-efficients
(see, e.g., \cite{boaris,boaris1,abdul} and references therein). In the present work, we will be interested in a system of two coupled Gross-Pitaevskii equations with a coherent coupling between both components \cite{rabi,abad} along with time-varying scattering lengths and
external potential. \\

On the other hand, study of nonlinear waves like, protean rogue waves and breathers in BECs is also one of the main objectives of modern day research \cite{pana,bludov,bec_brea}. A formal mathematical description of single rogue wave is provided by the  nonlinear Schr\"odinger (NLS) equation in the self-focusing regime \cite{solli}. Here the mechanism which leads to the generation of rogue waves is nonlinear wave mixing, that generates modulation instability (MI) of the continuous wave (CW) background~\cite{mi}. The nonlinear development of MI is described by families of exact breathers. A special member of this solution family is the Peregrine soliton \cite{peregrine}, which represents a wave that is localized both in space and time dimensions. In recent years, rogue waves and breathers have attracted much attention \cite{akmbec,spinrogue,bec_brea,degas}. A number of recent reviews have attempted to summarize the study of rogue waves in different contexts \cite{review}. The space-periodic breather type solution of NLS system has been obtained by Akhmediev et al., \cite{ussr} and the time-periodic solution of NLS was derived by Ma \cite{ma}.
\\

In this paper, we carry out a thorough analysis of the dynamics of superposed nonlinear waves
in autonomous and non-autonomous coherently coupled Gross-Pitaevskii (CCGP) system. The autonomous CCGP equations considered in this paper are shown to be integrable in Ref.~\cite{Park} and Kanna et al., have constructed the explicit soliton solutions of the underlying equation by applying a non-standard Hirota's bilinearization procedure in Ref. \cite{kannajpa}. Very recently, in Ref.~\cite{zhao}
 by applying the  Darboux transformation method, the soliton and rogue wave solutions of CCGP equations
 have been constructed. In the present work, we decouple the autonomous CCGP equations into two uncoupled NLS equations and make use of their different nonlinear wave solutions profitably to construct the superposed nonlinear wave solution of the autonomous CCGP equations and study their subsequent dynamics. Then we make use of these superposed nonlinear waves to obtain the nonlinear wave solutions of the non-autonomous CCGP
   equations with the aid of a similarity transformation. We have observed an interesting soliton phenomenon namely, soliton compression and also shown the existence of creeping soliton. The obtained solutions are new and display rich dynamical features that can find applications in atom interferometry and matter wave switches.\\

The remaining part of this paper is arranged as follows: We present the model equation and its physical significance in Sec. II. In Sec.
 III, the set of autonomous CCGP equations is converted into two independent NLS systems.
 We review the solutions of scalar (decoupled) NLS system and discuss their properties in Sec. III A.
 Sec. III B, deals with the interesting coherent structures of superposed nonlinear waves,
 namely colliding Ma breathers, coexisting rogue wave and Ma breathers, and coexisting
 Ma and Akhmediev breathers. In Sec. IV, we convert the non-autonomous CCGP system into an integrable autonomous
 CCGP system by using a similarity transformation. Also we clearly bring out the role of variable
 nonlinearity by considering two types of time-varying nonlinearities namely, kink-like and
 periodic modulated (Mathieu function) nonlinearity.
   Finally, the results are summarized in Sec.~V.\\

\section{The Model}

Here we consider the quasi-one-dimensional (cigar-shaped) BEC. For this case, the three dimensional CCGP system can be reduced to an one dimensional system and we can write the governing equations in dimensionless form as \cite{rabi,abad}
\bes\bea
i\frac{\partial\psi_1}{\partial t}&=&L_1\psi_1-\alpha_1(|\psi_1|^2+2|\psi_2|^2)\psi_1-\alpha_2 \psi_2^2\psi_1^*,\\
i\frac{\partial\psi_2}{\partial t}&=&L_2\psi_2-\alpha_1(2|\psi_1|^2+|\psi_2|^2)\psi_2-\alpha_2 \psi_1^2\psi_2^*,
\eea\label{nccnls}\ees
where $L_i$ = $-\partial_x^2+U_i(x,t)$ with $i=1,2$. Here, we measure
 the length and energy in units of
$a_{ho} = \sqrt{\hbar/m\omega_\bot}$, where $a_{ho}$ is the characteristic length of the
condensate and $\hbar\omega_\bot$, in which $\omega_\bot$ is the transversal frequency.
 In Eqs.~(\ref{nccnls}), $\psi_{j}$~($j=1,2$) are the condensate wave functions;
 the coefficients $\alpha_1$ and $\alpha_2$ introduce incoherent and coherent coupling between the two components and
  they can be tuned using Feshbach resonance mechanism and $U_i(x,t)(i=1,2)$ are the external potentials. For the homogeneous system the external potentials $U_i(x,t)$= 0. \\

 The physical significance of the above proposed model can be realized in the following two contexts. (i) In spin-1 BECs, the governing equation is three-component GP equations with the
components [i.e., $\Psi$=($\psi_{+1}/\sqrt{2}$, $\psi_0$, $\psi_{-1})^T$] corresponding to the three values of the vertical
spin projection, $m_F$ = $-1, 0,+1$ (see Eqs. (4) and (5) in ref. \cite{spincomplex}). For the special case ($\psi_{+1}=\psi_{-1}$ and $c_0=c_2=-c$), the set of three-component GP equations is reduced to two component GP system (\ref{nccnls}) with $\alpha_1$=$\alpha_2$=$c$. (ii) In the context of nonlinear optics, the model equation exactly same as that of (\ref{nccnls}) governs the pico-second pulse propagation in nonlinear Kerr media with low  birefringence or beam propagation in weakly anisotropic media and is usually referred as coherently coupled nonlinear Schr\"odinger equations \cite{book,kannajpa}.

Apart from these, Eqs.~(\ref{nccnls}) can also be viewed as the continuum limit of the model for the discrete coupled atomic condensates describing the four-wave mixing effects discussed in ref. \cite{discrete}

\section{Autonomous CCGP system}\label{trans}
First, we consider the homogeneous binary condensates ($U_i(x,t)=0$) with constant nonlinearity co-efficient $\alpha_1 = \alpha_2= \alpha$ in Eq.~(\ref{nccnls}). Here
$\alpha$ is a real integer and is independent of time. The resulting equations can be expressed as
\bes\bea
i\frac{\partial\psi_1}{\partial t}&=&-\frac{\partial^2\psi_1}{\partial x^2 }-\alpha(|\psi_1|^2+2|\psi_2|^2)\psi_1-\alpha\psi_2^2\psi_1^*,\\
i\frac{\partial\psi_2}{\partial t}&=&-\frac{\partial^2\psi_2}{\partial x^2 }-\alpha(2|\psi_1|^2+|\psi_2|^2)\psi_2- \alpha\psi_1^2\psi_2^*.
\eea\label{ccnls}\ees
 In this paper, we refer to Eqs.~(\ref{ccnls}) as autonomous CCGP equations.
These autonomous CCGP equations are integrable \cite{Park} and several interesting soliton solutions
have been obtained in Ref.\cite{kannajpa} by applying a non-standard Hirota's bilinearization technique.
We note that Eqs.~(\ref{ccnls})
 with $\alpha=2$, can be decoupled into two independent NLS equations of the form
\bea
~~~~~~~~~~iu_{j,t}+u_{j,xx} + 2u_j^*u_j^2=0,\qquad j=1,2,
\label{singlenls}
\eea
by applying a linear transformation
\bes\bea
&&\psi_1(x,t) = \frac{1}{2}(u_1+u_2),\\ &&\psi_2(x,t) = \frac{1}{2}(u_1-u_2),
\eea\label{q1}
\ees
where $u_1$ and $u_2$ are arbitrary analytic functions of $x$ and $t$ and satisfy the NLS equation (\ref{singlenls}).

The above NLS system (\ref{singlenls}) is completely integrable
and has been investigated extensively by using various analytical methods.
Interestingly, it admits many kinds of nonlinear wave solutions such as solitons,
rogue waves and breathers (including Ma and Akhmediev breathers) \cite{Chab,tkopt}
 and they have been experimentally observed under different contexts. Particularly the Peregrine soliton of
 the NLS system has been recently observed experimentally in nonlinear optics \cite{kibler}.

In this paper, we address the dynamical features of
superposition of these various structures which are also possible solutions of Eqs.~(\ref{ccnls}).
As a prelude we briefly revisit some of the interesting nonlinear wave solutions of the scalar NLS system (\ref{singlenls}) in the following sub-section.

\subsection{Revisit of nonlinear waves in the NLS system }\label{trans}
\subsection*{(i)~ Bright one-soliton}
The bright one-soliton solution $u_1$ of the NLS equation (\ref{singlenls}) with $j=1$
is obtained as \cite{bookml}
\bea
u_1(x,t)  = k^{(1)}_{R}~\mbox{sech}(z^{(1)}_{2})e^{iz^{(1)}_{1}},
\label{soliton}
\eea
where $z^{(1)}_{1} = k^{(1)}_{I}x+((k^{(1)}_{R})^2-(k^{(1)}_{I})^2)t+\eta_{I}^{(1)}$, $z^{(1)}_{2} =
k^{(1)}_{R}(x-2k^{(1)}_{I}t)+\eta_{R}^{(1)}+R^{(1)}$, and
$R^{(1)} = \ln \left[\frac{1}{2k^{(1)}_{R}}\right]$.
The above one-soliton solution $u_1$ is characterized by four real parameters $k^{(1)}_{R}$, $k^{(1)}_{I}$,
 $\eta_{R}^{(1)}$ and $\eta_{I}^{(1)}$. Similar solution for $u_2$ can be obtained by replacing the superscript (1) by (2). Here and in the following $R$ and $I$ appearing in the suffices of the soliton parameters denote the real and imaginary parts, respectively. The amplitude and velocity of the NLS soliton can be controlled by tuning $k^{(1)}_{R}$ and $k^{(1)}_{I}$, respectively,
 while the soliton position depends on the parameter $\eta_{R}^{(1)}$. %and pha which are instrumental in defining the nature of  affect the soliton amplitudes, velocity, and soliton position respectively.
%\begin{figure}[H]
%\centering\includegraphics[width=0.4\linewidth]{Fig_1}
%\caption{Evolution of bright one-soliton of the NLS equation (\ref{singlenls}) for $k^{(1)}_{R}=1.3$, $k^{(1)}_{I}=0$, $\eta_{R}^{(1)}=1$ and  $\eta_{I}^{(1)}=0.2$.}
%\label{solfig}
%\end{figure}
%Such a bright one-soliton propagation in NLS equation is shown in Fig.~\ref{solfig}. Similar soliton will result for $u_2$ also.%depicts the density plot of one soliton.\\

\subsection*{(ii)~ Bright two-soliton}
The bright two-soliton solutions of the above two different NLS systems (\ref{singlenls}) ($u_j$ with $j=1$ corresponds to the
solution of the first NLS system and $j=2$ corresponds to the
solution of the second NLS system) are given by \cite{bookml}
\bes\bea
&&u_j(x,t) =\frac{G^{(j)}}{F^{(j)}}, \quad j=1,2,
\label{twosoliton}
\eea
where
\bea &&G^{(j)}=e^{\eta^{(j)}_{1}}+e^{\eta^{(j)}_{2}}+e^{\eta^{(j)}_{1}+\eta_{1}^{(j)*}+\eta^{(j)}_{2}+\delta^{(j)}_{1}}+e^{\eta^{(j)}_{1}+\eta_{2}^{(j)*}+
\eta^{(j)}_{2}+\delta^{(j)}_{2}},\\
&&F^{(j)}=1+e^{\eta^{(j)}_{1}+\eta_{1}^{(j)*}+R^{(j)}_{1}}+e^{\eta^{(j)}_{1}+\eta_{2}^{(j)*}+\delta^{(j)}_{0}}+
e^{\eta^{(j)}_{2}+\eta_{1}^{(j)*}+\delta_{0}^{(j)*}}+e^{\eta^{(j)}_{2}+\eta_{2}^{(j)*}+R^{(j)}_{2}}+
e^{\eta^{(j)}_{1}+\eta_{1}^{(j)*}+\eta^{(j)}_{2}+\eta_{2}^{(j)*}+R^{(j)}_{3}}.
\eea \ees
In the above solution, $\eta^{(j)}_{1}=k^{(j)}_{1} x+i (k_{1}^{(j)})^2t+\eta_{1}^{(j)(0)}$, $\eta^{(j)}_{2}=
k^{(j)}_{2} x+i (k_{2}^{(j)})^2t+\eta_{2}^{(j)(0)}$, $e^{R^{(j)}_{1}}=\frac{1}{(k^{(j)}_{1} + k_{1}^{(j)*})^2}$,
$e^{R^{(j)}_{2}}=\frac{1}{(k^{(j)}_{2} + k_{2}^{(j)*})^2}$, $e^{\delta^{(j)}_{0}}=\frac{1}{(k^{(j)}_{1}+k_{2}^{(j)*})^2}$,
$e^{\delta_{0}^{(j)*}}=\frac{1}{(k_{1}^{(j)*}+k^{(j)}_{2})^2}$, $e^{\delta^{(j)}_{1}}=\frac{(k^{(j)}_{1}-k^{(j)}_{2})^2}{(k^{(j)}_{1}+k_{1}^{(j)*})^2(k_{1}^{(j)*}+k^{(j)}_{2})^2}$,
 $e^{\delta^{(j)}_{2}}=\frac{(k^{(j)}_{2}-k^{(j)}_{1})^2}{(k^{(j)}_{1}+k_{2}^{(j)*})^2(k_{2}^{(j)*}+k^{(j)}_{2})^2}$ and
 $e^{R^{(j)}_{3}}=\frac{(k^{(j)}_{1}-k^{(j)}_{2})^2(k_{1}^{(j)*}-k_{2}^{(j)*})^2}{(k^{(j)}_{1}+k_{1}^{(j)*})^2(k_{1}^{(j)*}+k^{(j)}_{2})^2(k^{(j)}_{1}+k_{2}^{(j)*})^2(k^{(j)}_{2}+k_{2}^{(j)*})^2}$.

  The bright two-soliton solution of the NLS equation is characterized by four complex parameters
  $k^{(j)}_{1}$, $k^{(j)}_{2}$, $\eta_{1}^{(j)(0)}$ and $\eta_{2}^{(j)(0)}$.
  Here the real parts of $k^{(j)}_{1}$ and $k^{(j)}_{2}$ control the amplitude of the respective soliton,
  while their imaginary parts influence the velocity. One can observe two kinds of dynamics from the
  above two-soliton solution: two-soliton bound state with beating for same soliton velocities ($k^{(1)}_{1I}=k^{(1)}_{2I}$)
   and a head-on elastic soliton collision for equal but opposite soliton velocities ($k^{(1)}_{1I}=-k^{(1)}_{2I}$).
  % In Fig.~\ref{twosolfig}, we have shown such a two-soliton bound state (left panel) and elastic collision
 %  (right panel) of NLS equation (\ref{singlenls}).
%\begin{figure}[H]
%\centering\includegraphics[width=0.4\linewidth]{Fig_2a}~~\includegraphics[width=0.4\linewidth]{Fig_2b}
%\caption{Bound solitons (left panel) for $k^{(1)}_{1}=1.25-0.2i$, $k^{(1)}_{2}=1-0.2i$ and $\eta^{(1)(0)}_1$ =$\eta^{(1)(0)}_2$=1; Head on collision of bright two solitons
%(right panel) for $k^{(1)}_{1}=1.25+0.2i$, $k^{(1)}_{2}=1-0.2i$, and $\eta^{(1)(0)}_1$ %=$\eta^{(1)(0)}_2$=1
%in system(\ref{singlenls}).}
%\label{twosolfig}
%\end{figure}
%Similar interactions also occur for $u_2$.
\subsection*{(iii)~ Rogue wave}
Rogue waves are nothing but a sudden peak (hump) localized both in space and time with maximum amplitude on
 a continuous wave background.
 The rogue wave (rational) solutions ($u_j,~j=1,2$) of the two independent NLS systems (\ref{singlenls})
 obtained in Ref.~\cite{rogue}
 can be written as
\bea
u_j(x,t) = \left[1-\frac{2(1+4i|\delta^{(j)}|^2t)}{2|\delta^{(j)}|^2(x^2+4|\delta^{(j)}|^2t^2+\frac{1}{4|\delta^{(j)}|^2})}\right]\times \delta^{(j)} e^{2i|\delta^{(j)}|^2 t}, \quad j=1,2.
\label{rogue}
\eea
Here the arbitrary complex parameter $\delta^{(j)}$ influences the
 amplitude as well as the width of the rogue wave in the $j^{th}$ NLS system. In general, the amplitude
 of rogue wave is three-times (or more) higher than the background carrier wave.

%\begin{figure}[H]
%\centering\includegraphics[width=0.5\linewidth]{Fig_3}%
%\caption{Dynamics of Rogue wave for $\alpha^{(1)}=\sqrt{1.3}$.}
%\label{rogfig}
%\end{figure}

\subsection*{(iv)~ Akhmediev breather}
Akhmediev breather (AB) solution ($u_j$, where $j=1$ corresponds to the
solution of the first NLS system and $j=2$ corresponds to the
solution of the second NLS system) of the two independent NLS equations (\ref{singlenls}) is given by
\bea
u_j(x,t) = \left[\frac{\cosh(\Omega t-2ia^{(j)})-\cos(a^{(j)})\cos(b^{(j)}x)}{\cosh(\Omega t)-\cos(a^{(j)})\cos(b^{(j)}x)}\right]\times e^{2it}, \quad j=1,2,
\label{akh}
\eea
where $\Omega = 2 \sin(2a^{(j)})$, $b^{(j)} = 2 \sin(a^{(j)})$ and $a^{(j)}$ $\in$ $\mathbb{R}$.
%\begin{figure}[H]
%\centering\includegraphics[width=0.5\linewidth]{Fig_4}
%\caption{Evolution of Akhmediev breather for $a^{(1)}=1$.}
%\label{Ak-breather}
%\end{figure}
The role of parameter $a^{(j)}$ in AB is directly related to the
number of breathers and inversely related to its maximum amplitude.
 To be more precise, for increasing value of $a^{(j)}$ the number of breathers over a definite region
 increases but their amplitude decreases, whereas the reverse scenario takes place for decreasing value of $a^{(j)}$ \cite{Chab}.

\subsection*{(v)~ Ma breather}
The Ma breather (MB) solutions of the two independent NLS equations (\ref{singlenls}) can be expressed as
\bea
u_j(x,t) = \left[\frac{\cos(\Omega t-2ia^{(j)})-\cosh(a^{(j)})\cosh(b^{(j)}x)}{\cos(\Omega t)-\cosh(a^{(j)})\cosh(b^{(j)}x)}\right]\times e^{2it}, \quad j=1,2,
\label{Ma}
\eea
where $\Omega = 2 \sinh(2a^{(j)})$, $b^{(j)} = 2 \sinh(a^{(j)})$ and $a^{(j)}$ $\in$ $\mathbb{R}$.
%\begin{figure}[H]
%\centering\includegraphics[width=0.5\linewidth]{Fig_5}
%\caption{Evolution of Ma breather for $a^{(1)}=0.2$.}
%\label{Ma-breather}
%\end{figure}
These Ma breathers are localized in spatial~($x$) coordinate and periodic in time~($t$) coordinate.
The effect of parameter $a^{(j)}$ is straightforward in the case of MB, where the number of
breathers as well as their maximum amplitude increases (decreases) when $a^{(j)}$ is increased (decreased) \cite{Chab}.

\subsection{Dynamics of superposed nonlinear waves}
In the starting of this section, we have shown that the solution of the autonomous CCGP system (\ref{ccnls}) can be obtained by superposing the solutions of two different NLS systems. Following that, we revisit several nonlinear wave solutions of NLS system.

Now we look into the various combinations of these solutions discussed in Sec.
IIA and construct several superposed solutions of the autonomous CCGP system. This will give
rise to the possibility of non-trivial localized nonlinear coherent structures in the autonomous CCGP system.\\

\noindent{\textbf{Case~(i)} Superposition of two different one-soliton solutions}

To start with, we consider two different one-soliton solution forms obtained from (\ref{soliton}) with
$j=1$ and $2$ and write down the solutions $\psi_j,(j=1,2)$ of the autonomous CCGP system using (\ref{q1}).

\begin{figure}[!pht]
\centering\includegraphics[width=0.8\linewidth]{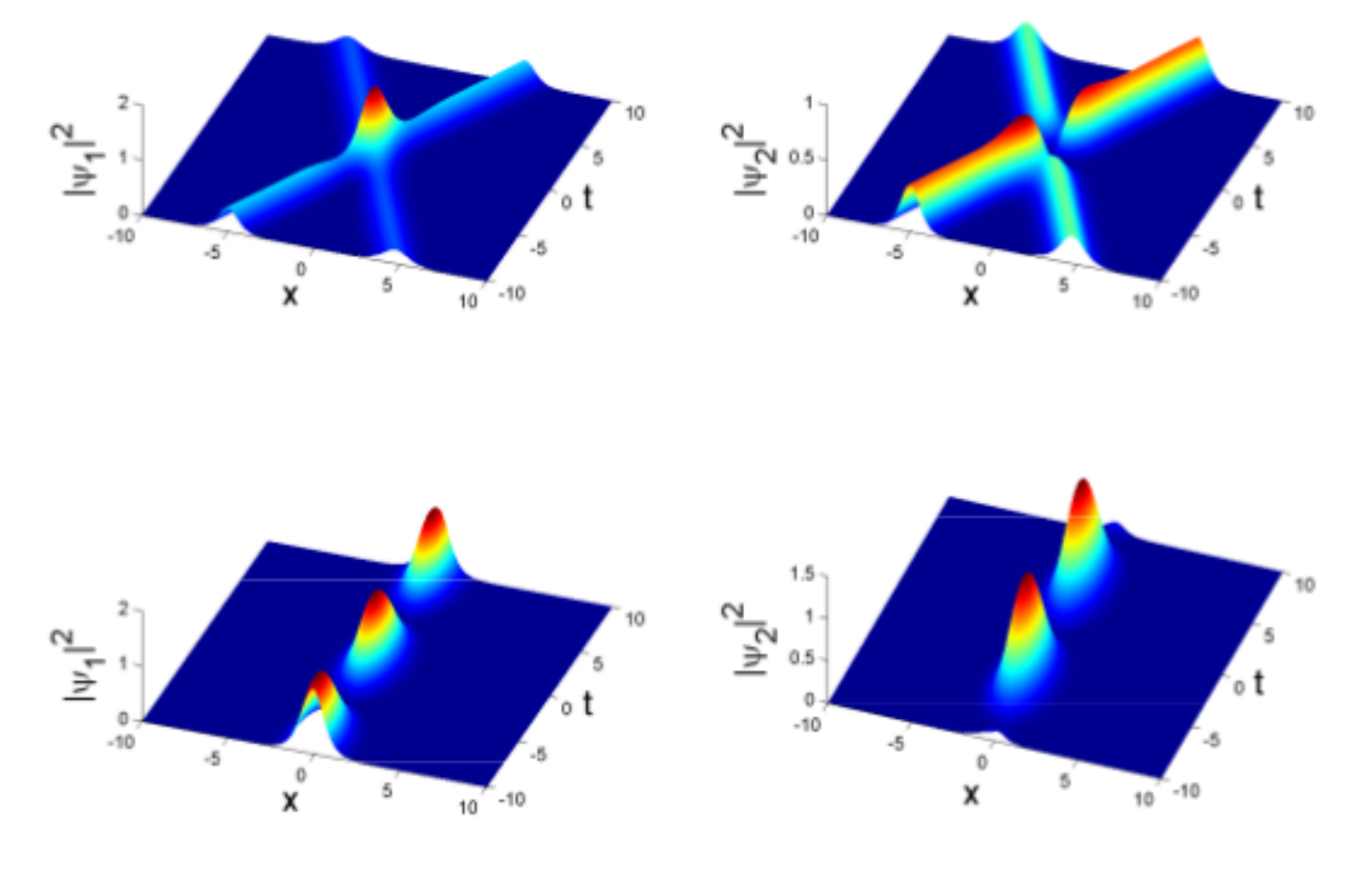}
\caption{Superposition of two different one-soliton. Top panel: Soliton collision with parameters
$k^{(1)}_R=1.3$, $k^{(1)}_{I}=0.25$, $k^{(2)}_{R}=1$, $k^{(2)}_{I}=-0.25$, $\eta_{R}^{(1)(0)}=\eta_{R}^{(2)(0)}=1$ and $\eta_{I}^{(1)(0)}=\eta_{I}^{(2)(0)}=0.2$;
Bottom panel: Bound soliton for the parametric choice $k^{(1)}_{R}=1.3$, $k^{(1)}_{I}=0$, $k^{(2)}_{R}=1$,
 $\eta_{R}^{(1)(0)}=\eta_{R}^{(2)(0)}=1$ and $\eta_{I}^{(1)(0)}=\eta_{I}^{(2)(0)}=0.2$.}
\label{s-s}
\end{figure}

First we superimpose two one-solitons travelling with different amplitudes and opposite velocities.
The resultant structure is a head on collision of two soliton in the $\psi_1$ and $\psi_2$ components. This is depicted in Fig.~\ref{s-s}.

The expression for $|\psi_1|^2 = \frac{1}{4} \left[(k^{(1)}_{R})^2 \mbox{sech}^2(z^{(1)}_{2})+(k^{(2)}_{R})^2
\mbox{sech}^2(z^{(2)}_{2})+2 k^{(1)}_{R}k^{(2)}_{R}\mbox{sech}(z^{(1)}_{2})\mbox{sech}(z^{(2)}_{2})
\cos(z^{(1)}_{1}-z^{(2)}_{1})\right]$ and $|\psi_2|^2 = \frac{1}{4} \left[(k^{(1)}_{R})^2
\mbox{sech}^2(z^{(1)}_{2})+(k^{(2)}_{R})^2 \mbox{sech}^2(z^{(2)}_{2})-2 k^{(1)}_{R}k^{(2)}_{R}
\mbox{sech}(z^{(1)}_{2})\mbox{sech}(z^{(2)}_{2})\cos(z^{(1)}_{1}-z^{(2)}_{1})\right]$.
At the origin [$(x,t)$=0], the intensity of the resulting superposed coherent structure attains a maximum value in the $\psi_1$ component while it reaches a minimum value in the $\psi_2$ component. This is a consequence of conservation of total energy $(\int (|\psi_|^2+|\psi_2|^2) dt = const.)$. If we choose $k^{(1)}_{I}$ and $k^{(2)}_{I}$ to be zero or to be equal with $z^{(1)}_{1} \neq z^{(2)}_{1}$,
 we obtain soliton bound state with oscillations along $t$ axis. This is clearly shown in the bottom panel
 of Fig.~\ref{s-s}. Another interesting possibility arises for $k^{(1)}$=$k^{(2)}$, where the superposed
 soliton exists only in the $\psi_1$ component and is absent in the $\psi_2$ component. Thus we can engineer the nature of soliton propagation by tuning the $k^{(j)}$ parameters appropriately.\\

\noindent{\textbf{Case~(ii)} Superposition of one-soliton ($u_1$) with two-soliton ($u_2$)}

\begin{figure}[!pht]
\vspace{-0.5cm}
\centering\includegraphics[width=0.8\linewidth]{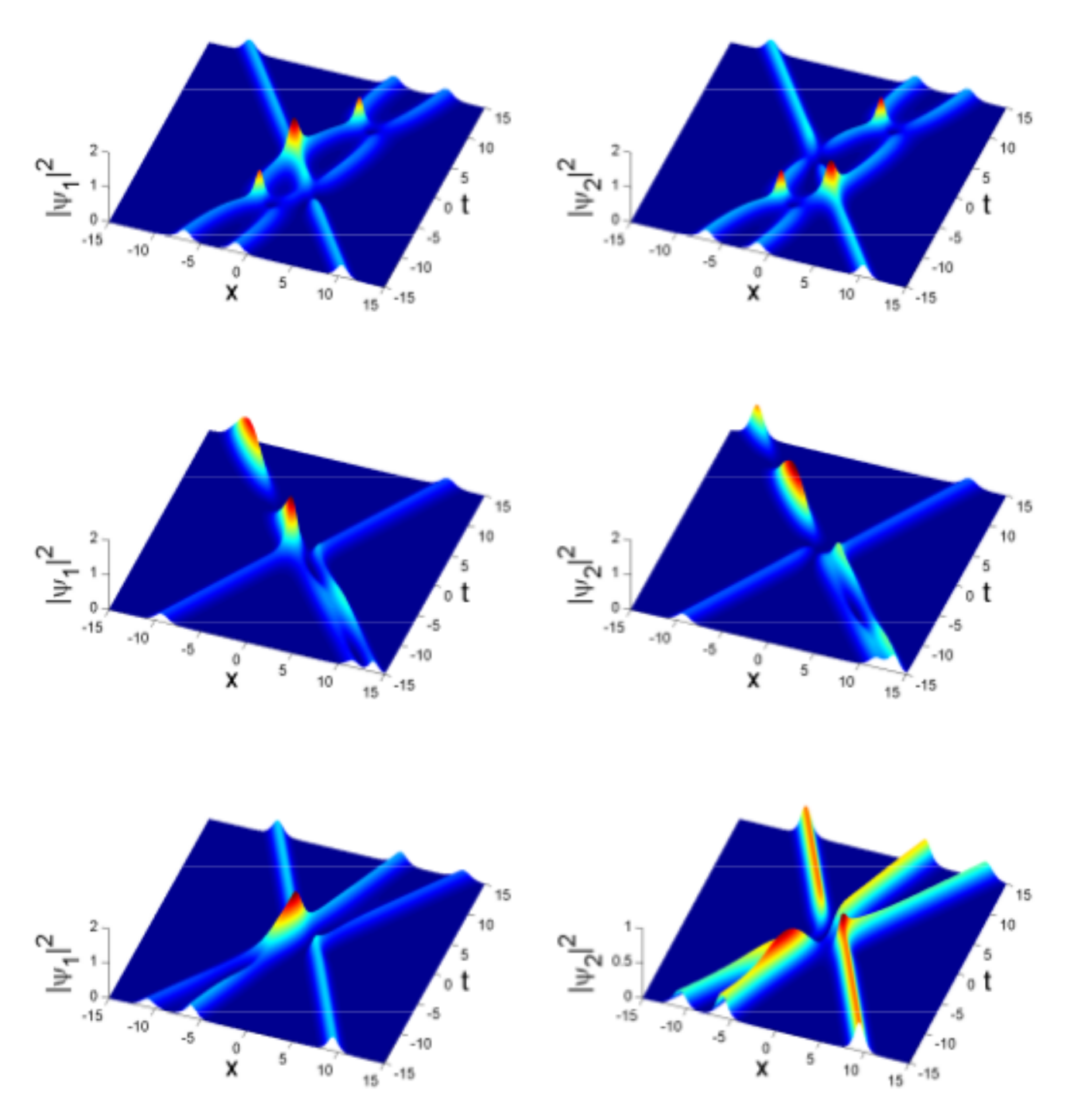}
\caption{Superposition of single- and two- solitons. Top Panel:
 Collision of bound state and soliton with parameters $k^{(1)}=1.2-0.35i$, $k^{(2)}_{1}=1.2+0.2i$, $k^{(2)}_{2}=1+0.2i$,
 $\eta^{(2)(0)}_{1}=1$, $\eta^{(2)(0)}_{2}=1$, $\eta^{(1)}_R=1$, and $\eta^{(1)}_I=0.2$;
Middle Panel: Collision of one- soliton with bound soliton with parameters $k^{(1)}=1-0.4i$, $k^{(2)}_{1}=1.25-0.4i$, $k^{(2)}_{2}=1+0.3i$,
 $\eta^{(2)(0)}_{1}=1$, $\eta^{(2)(0)}_{2}=1$, $\eta^{(1)}_R=1$, and $\eta^{(1)}_I=0.1$; Bottom Panel: Three soliton collision with parameters $k^{(1)}=1.2+0.2i$, $k^{(2)}_{1}=1.3-0.25i$,
 $k^{(2)}_{2}=1+0.35i$ and all other parameters are same as given in the top panel.}
\label{fig7}
\end{figure}

Next we construct $\psi_1$ and $\psi_2$ (see Eq.~(\ref{q1})) by superposing
one-soliton solution ($u_1$) with two-soliton solution ($u_2$).
Fig.~\ref{fig7} depicts the density plot of such superposed nonlinear coherent structures.
Here we discuss two possible interesting coherent structures in the $\psi_1$ and $\psi_2$ components:
(i)~superposition of bound soliton (which can be obtained by setting same values for the imaginary parts
of $k^{(2)}_{1}$ and $k^{(2)}_{2}$ parameters in the two-soliton solution) with one soliton.
This is shown in the top panel of Fig.~\ref{fig7}. There is another possibility which is shown in the middle panel of figure (\ref{fig7}). Here the coherent structure displays a collision of bound soliton with a single soliton. Specifically a double-hump breather is converted into a single-hump breather with increased intensity after collision. (ii)~superposition of two-soliton with one-soliton in which all the three solitons are travelling at different velocities.
 To obtain this we set different values for $k^{(1)}$, $k^{(2)}_{1}$ and $k^{(2)}_{2}$ (see bottom panel of Fig.~\ref{fig7}). The resulting coherent structure is an interaction of three independent solitons.

\noindent{\textbf{Case~(iii)} Superposition of one-soliton ($u_1$) with Akhmediev breather ($u_2$)}

In this case, we choose the one-soliton solution for $u_1$ as given by Eq.~(\ref{soliton}) and
Akhmediev breather (spatially periodic) form for $u_2$ in (\ref{q1}). After superposing a stationary soliton ($k^{(1)}_I=0$) with Akhmediev breather, we obtain special nonlinear coherent structures in which the Ma and
Akhmediev breathers coexist in both $\psi_1$ and $\psi_2$ components.
\begin{figure}[!pht]
\centering\includegraphics[width=0.8\linewidth]{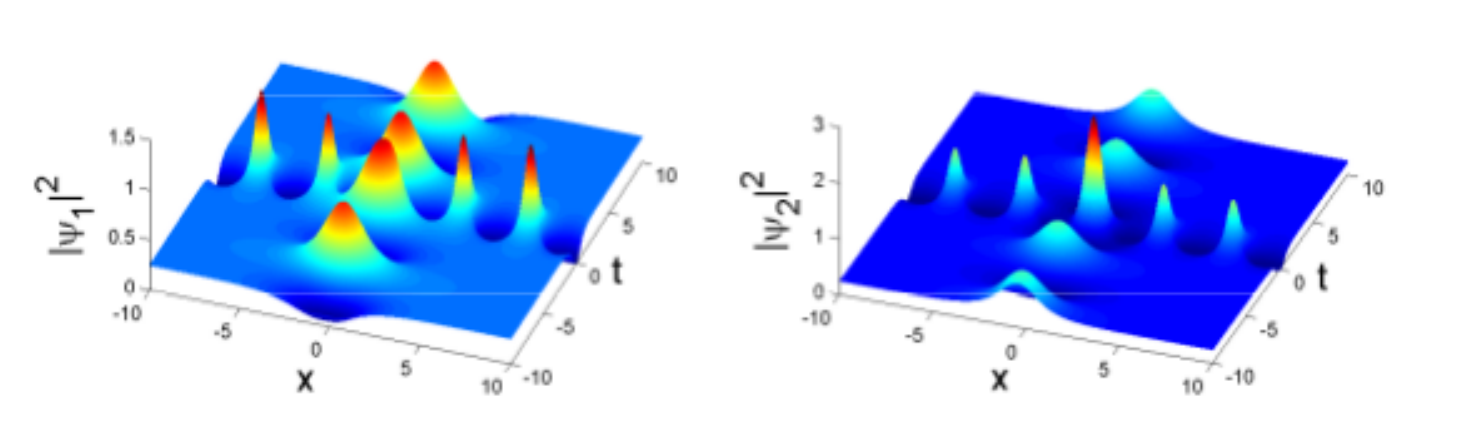}
\caption{Superposition of one-soliton with Akhmediev breather. The parameters $k^{(1)}_{R}=0.9$,
$k^{(1)}_{I}=0$, $a^{(2)}=1$, $\eta_{R}^{(1)}$=1, and $\eta_{I}^{(1)}$=0.2.}
\label{fig9}
\end{figure}
Such coherent structures are shown in Fig.~\ref{fig9}. For solitons with non-zero velocity, also similar co-existence of Ma and Akhmediev breathers exists.\\

\noindent{\textbf{Case~(iv)} Superposition of one-soliton ($u_1$) with Ma breather ($u_2$)}

\begin{figure}[!pht]
\centering\includegraphics[width=0.8\linewidth]{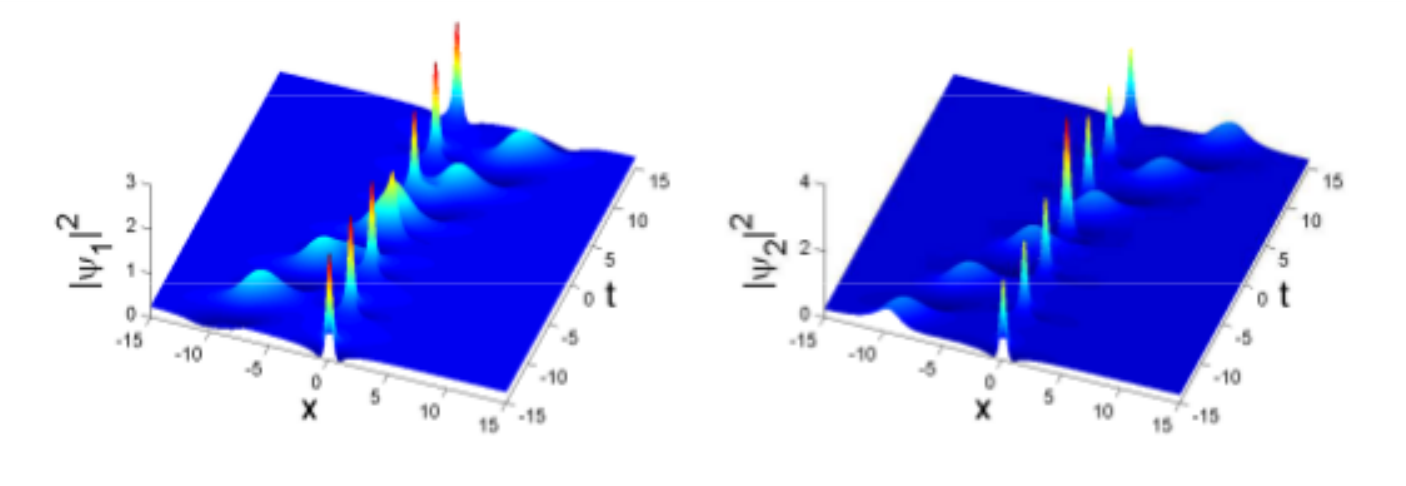}
\caption{Superposition of one-soliton with Ma breather. The parameters are $k^{(1)}_{R}=0.9$, $k^{(1)}_{I}=0.3$, $a^{(2)}=0.2$,
 $\eta_{R}^{(1)}=-1$, and $\eta_{I}^{(1)}=0.2$.}
\label{fig10}
\end{figure}

In this case, we choose the one-soliton solution as given by Eq.~(\ref{soliton}) and
MB (time periodic) form for $u_2$ in Eq.~(\ref{q1}). In $\psi_1$ and $\psi_2$ components,
we get nonlinear coherent structures akin to the collision of two MBs.
This is shown in Fig.~\ref{fig10} for illustrative purpose. If we choose the soliton velocity to be zero, the two MBs overlap with each other.

\begin{figure}[!pht]
\centering\includegraphics[width=0.8\linewidth]{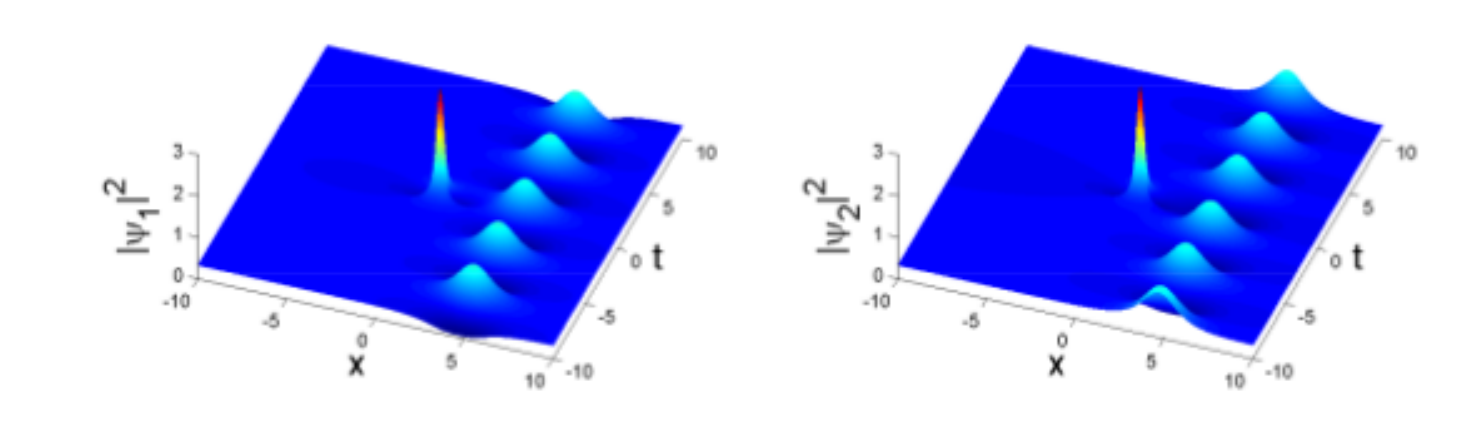}
\caption{Superposition of one-soliton with rogue wave with parameters $k^{(1)}_{R}=1$, $k^{(1)}_{I}=0$, $\eta_{R}^{(1)}=-4$, $\eta_{I}^{(2)}=0.2$ and $\delta^{(2)}$=$\sqrt{1.3}$.}
\label{fig8}
\end{figure}

\noindent{\textbf{Case~(v)}~Superposition of one-soliton ($u_1$) with a Rogue wave ($u_2$)}\\
In this case, for $u_1$ and $u_2$ in (\ref{q1}), we choose one soliton and rogue wave solution, respectively.
When we superpose these two types of nonlinear waves using the transformation (\ref{q1}),
the resultant structure is the coexistence of MB and rogue wave in the $\psi_1$ and $\psi_2$ components.
This is depicted in figure \ref{fig8}.

Here we note that the soliton is converted into a MB.
This can be understood by writing down the detailed expression for $|\psi_1|^2$ and  $|\psi_2|^2$. We find $|\psi_1|^2$= $\frac{1}{4}\left(|\lambda|^2+(k_{R}^{(1)})^2\mbox{sech}^2(z^{(1)}_2)+k_{R}^{(1)}\mbox{sech}(z^{(1)}_2)[2\lambda_R\cos(\chi)-2\lambda_I\sin(\chi)]\right)$ and $|\psi_2|^2$=$\frac{1}{4}\left(|\lambda|^2+(k_{R}^{(1)})^2\mbox{sech}^2(z^{(1)}_2)+k_{R}^{(1)}\mbox{sech}(z^{(1)}_2)[-2\lambda_R\cos(\chi)+2\lambda_I\sin(\chi)]\right)$,
 where $\chi$ = $2|\delta^{(2)}|^2t-z_1^{(1)}$ and

 $\lambda$=$\delta^{(2)} \left[1-\frac{2(1+4i|\delta^{(2)}|^2t)}{2|\delta^{(2)}|^2(x^2+4|\delta^{(2)}|^2t^2+\frac{1}{4|\delta^{(2)}|^2})}\right]$.
  We observe that the interference terms for $|\psi_1|^2$ and $|\psi_1|^2$ involve $cosine$ and $sine$ functions.
  Ultimately, this results in breathing oscillations in the soliton intensity along the $t$ axis thereby converting
  it into a MB.\\

\noindent{\textbf{Case~(vi)} Superposition of two-soliton solution ($u_1$) with Rogue wave ($u_2$)}

\begin{figure}[!pht]
\centering\includegraphics[width=0.8\linewidth]{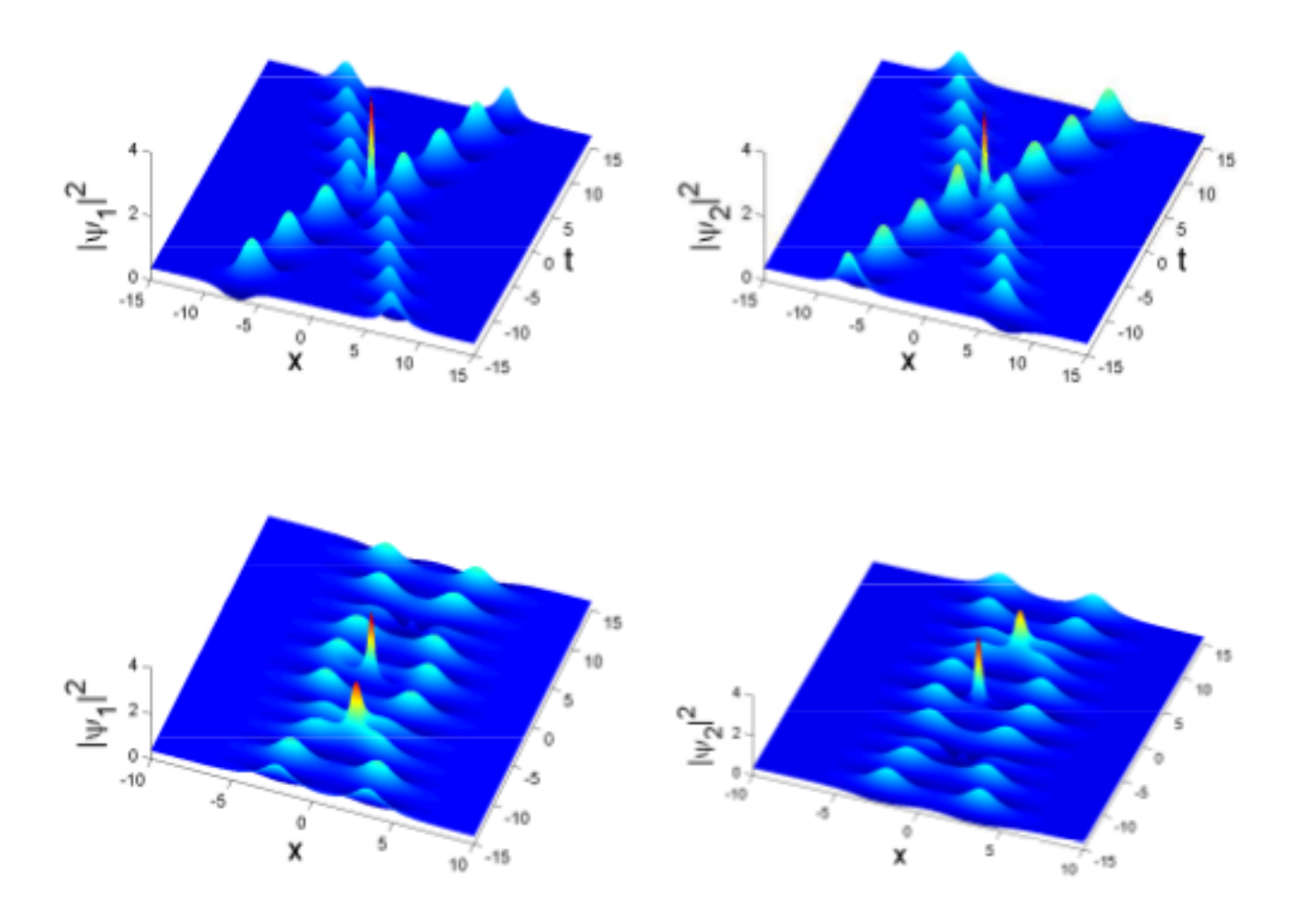}
\caption{Superposition of two-soliton with rogue wave.
Top Panel: Coexistence of interactions between two breathers and rogue wave with parameters are
$k^{(1)}_{1}=1.25+0.2i$, $k^{(1)}_{2}=1-0.2i$,
$\delta^{(2)}=\sqrt{1.5}$, and $\eta^{(1)(0)}_1$=$\eta^{(1)(0)}_2$=$2.5+i$.
Bottom panel: Breather bound states with parameters $k^{(1)}_{1}=1.2$, $k^{(1)}_{2}=1$, $\delta^{(2)}=\sqrt{1.5}$ and
$\eta^{(1)(0)}_1$=$\eta^{(1)(0)}_2$=$2.5+i$.}
\label{fig11}
\end{figure}

Here we choose two-soliton solution and rogue wave for $u_1$ and $u_2$, respectively, in Eq.~(\ref{q1}).
The two colliding solitons are now converted into interacting MBs. As a result of this we get a
nonlinear coherent structure comprising of a rogue wave and two colliding MBs.
This is shown in the top panel of Fig. \ref{fig11} for the choice of parameters
$k^{(1)}_{1}$ = $1 + 0.2i$, $k^{(1)}_{2}$ = $1- 0.2i$ and $\delta^{(2)}=\sqrt{1.5}$.
If we choose the imaginary part of $k^{(2)}_{1}$ and $k^{(2)}_{2}$ parameters to be zero,
which means the velocities of respective solitons are zero, then we obtain bound state with breathing oscillations (see bottom panel).

\noindent{\textbf{Case~(vii)} Superposition of Rogue waves}

\begin{figure}[!pht]
\centering\includegraphics[width=0.8\linewidth]{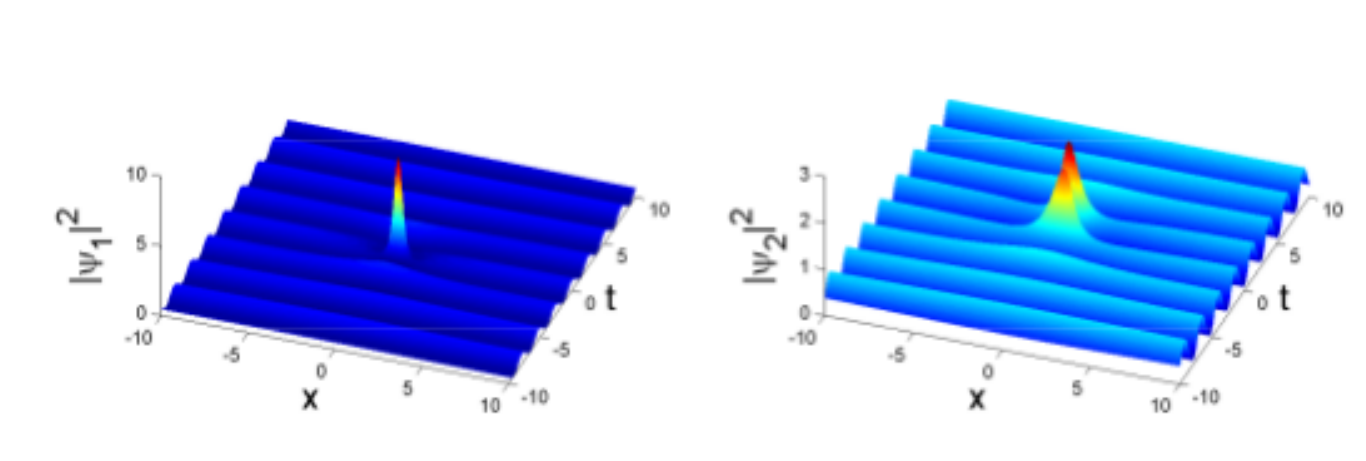}
\caption{Superposition of two different rogue waves. The parameters are
$\delta^{(1)}=\sqrt{1.5}$ and $\delta^{(2)}$=$\sqrt{0.3}-0.1i$.}
\label{fig12}
\end{figure}

We consider two distinct rogue wave solutions for $u_1$ and $u_2$, and insert them in Eq.~(\ref{q1}).
We plot the resulting intensities $|\psi_1|^2$ and $|\psi_2|^2$ in Fig.~\ref{fig12}.
In the $\psi_1$ component the rogue waves merge together along with a periodic oscillation of the background.
In the $\psi_2$ component we get a non-trivial twin peak rogue wave. Here too the background shows significant oscillations.
To facilitate the understanding of the oscillations we explicitly present the expressions
for $|\psi_1|^2$ and $|\psi_2|^2$.
$|\psi_1|^2=\frac{1}{4}\left(|\lambda_1|^2+|\lambda_2|^2+2A \cos[2(|\delta^{(1)}|^2-|\delta^{(2)}|^2)t]-2B\sin[2(|\delta^{(1)}|^2-|\delta^{(2)}|^2)t]\right)$
and \\
$|\psi_2|^2 = \frac{1}{4} \left( |\lambda_1|^2+|\lambda_2|^2-2A \cos[2(|\delta^{(1)}|^2-|\delta^{(2)}|^2)t]+
2B\sin[2(|\delta^{(1)}|^2-|\delta^{(2)}|^2)t]\right)$,
where $\lambda_1  = \delta^{(1)}  \left[1-\frac{2(1+4i|\delta^{(1)}|^2t)}{2|\delta^{(1)}|^2(x^2+4|\delta^{(1)}|^2t^2+\frac{1}{4|\delta^{(1)}|^2})}\right]$,
$\lambda_2  = \delta^{(2)}  \left[1-\frac{2(1+4i|\delta^{(2)}|^2t)}{2|\delta^{(2)}|^2(x^2+4|\delta^{(2)}|^2t^2+\frac{1}{4|\delta^{(2)}|^2})}\right]$,
$A$ and $B$ are real and imaginary parts of $\lambda_1\lambda_2^*$.
From this expression we find that the oscillations originate from the circular functions ($cosine$ and $sine$)
appearing in the cross terms and depend upon the difference $|\delta^{(1)}|^2-|\delta^{(2)}|^2$. Especially, for $\delta^{(1)}=\delta^{(2)}$,
the background oscillations are completely suppressed and the rogue wave appears only in the $\psi_1$ component while it disappears in the $\psi_2$ component.\\

\noindent{\textbf{Case~(viii)} Superposition of Rogue wave ($u_1$) with Akhmediev breather ($u_2$)}

\begin{figure}[!pht]
\centering\includegraphics[width=0.8\linewidth]{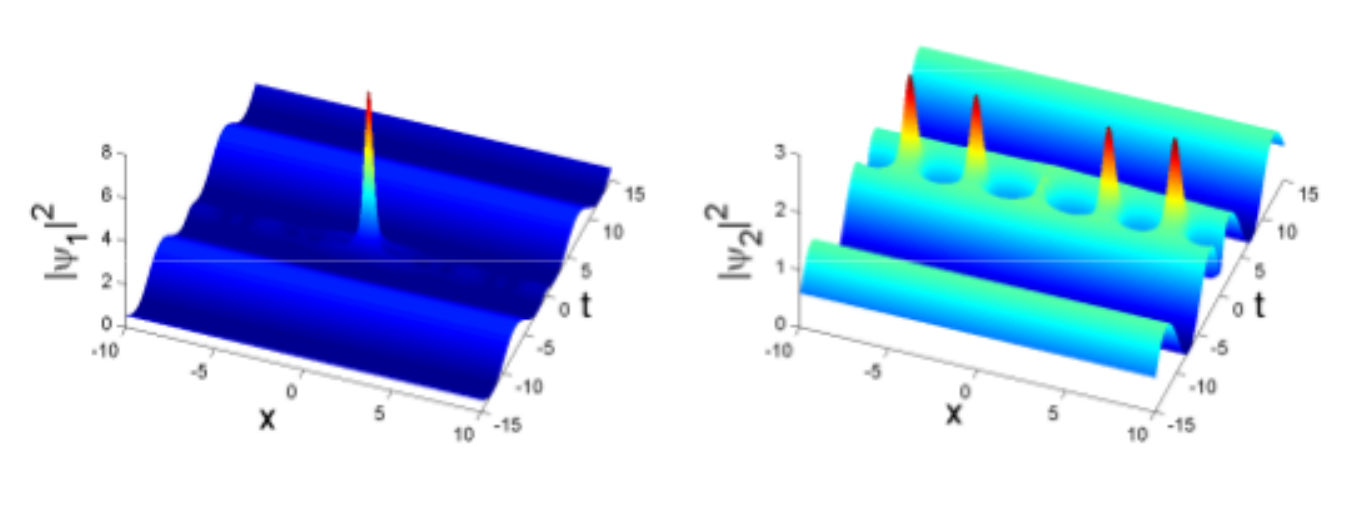}
\caption{Superposition of rogue wave with Akhmediev breather for $\delta^{(1)}=\sqrt{1.5}$ and $a^{(2)}=0.9$.}
\label{fig13}
\end{figure}

We choose the solution $u_1$ to admit rogue wave and $u_2$ to possess the form of Akhmediev breather.
 This type of superposition allows the possibility of interesting distinct nonlinear coherent
  structures in the $\psi_1$ and $\psi_2$ components.
  In the $\psi_1$ component the dominant behaviour is displayed by the rogue wave whose amplitude
  is increased significantly while that of the AB gets suppressed.
  The reverse scenario takes place in the $\psi_2$ component, that is,
  here the amplitude of the rogue wave is suppressed whereas that of the AB is enhanced significantly.
   This shows that there is an energy redistribution among the components.
  In both the components the background executes oscillations.
  Particularly in the $\psi_2$ component the background oscillates vibrantly. Such coherent structures are shown in Fig.~\ref{fig13}.

\noindent{\textbf{Case~(ix)} Superposition of Rogue wave ($u_1$) with Ma breather ($u_2$)}

\begin{figure}[!pht]
\centering\includegraphics[width=0.8\linewidth]{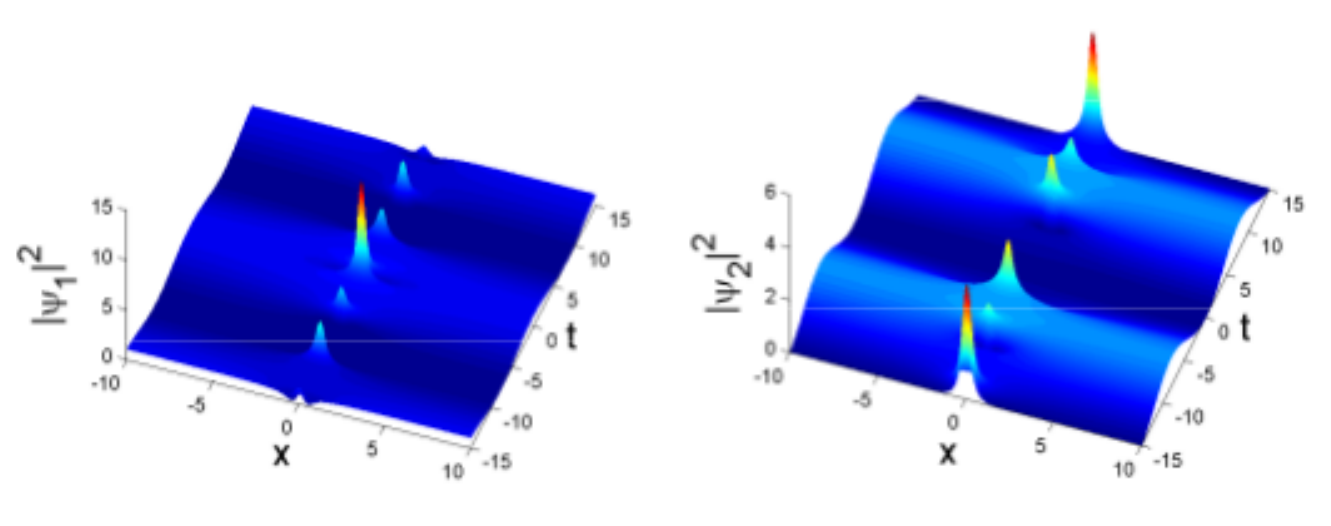}
\caption{Superposition of rogue wave with Ma breather.
The parameters are $\delta^{(1)}=\sqrt{1.2}$ and $a^{(2)}=0.3$}
\label{fig14}
\end{figure}

 Finally, we consider the rogue wave solution for $u_1$ and Ma breather solution for $u_2$ in Eq.~(\ref{q1}).
  The resulting intensity plots for the $\psi_1$ and $\psi_2$ components are shown in Fig.~\ref{fig14}.
  This type of superposition results in nonlinear coherent structures with special dynamical properties.
  In the $\psi_1$ component the rogue wave appears predominantly. But it disappears completely in the $\psi_2$ component.
  In both the components the MB periodically reaches maximum and minimum intensities.
  Especially in the $\psi_2$ component the breather is dominant.
  Thus here too an energy redistribution takes place among the components $\psi_1$ and $\psi_2$. Also,
  the background displays periodic oscillations in both the components.

In the following table, we have summarized all the above interesting nonlinear superposed coherent structures.\\
\begin{table}[H]
\caption {Superposition of several kinds of nonlinear waves}
\centering
\begin{tabular}{|c|c|p{2.1cm}|p{5.5cm}|p{5.5cm}|c|c|}
  \hline
             Case & Form of & ~~Form of & \multicolumn{2}{c|}{Nature of nonlinear coherent structures} \\
  \cline{4-5}
  &$~~u_1$&~~~~$u_2$& ~~~~~~~$\psi_1$ component & ~~~~~~~$\psi_2$ component \\
  \hline
  \hline
             (i) &  One-soliton & One-soliton & (a) Two soliton collision with higher intensity, (b) Oscillating bound soliton with larger period.& (a) Two soliton collision with lower intensity, (b) Oscillating bright soliton with smaller period.\\
  \hline
             (ii) & One-soliton & Two-soliton & (a) Collision of bound soliton with one soliton, (b) Three soliton collision with higher amplitude, and (c) Three soliton bound state. & (a) Similar behavior as in $\psi_1$ component, (b) Three soliton collision with lower amplitude, and (c) Three soliton bound state. \\
  \hline
             (iii) & One-soliton & Rogue wave & \multicolumn{2}{c|}{\multirow{1}{*}{Coexistence of Ma breather with rogue wave.}}\\
  \hline
             (iv) & One-soliton & Akhmediev breather & Coexistence of Ma and Akhmediev breathers with lower amplitude. & Coexistence of Ma and Akhmediev breathers with higher amplitude.\\
  \hline
             (v) & One-soliton & Ma breather & \multicolumn{2}{c|}{\multirow{1}{*}{Colliding Ma breathers}}.\\
  \hline
           (vi) & Two-soliton & Rogue wave & \multicolumn{2}{c|}{\multirow{1}{*}{(a) Coexistence of colliding  Ma breathers with rouge wave}}.\\
           &&& \multicolumn{2}{c|}{\multirow{1}{*}{\hspace{-1.3cm}(b) Bound state of Ma breathers with rouge wave}}.\\
  \hline
             (vii) & Rogue wave & Rogue wave & Higher amplitude rogue wave with oscillating background. &  Lower amplitude twin peak rogue waves with oscillating background.\\
 \hline
             (viii) & Rogue wave & Ma breather &  Rogue wave is dominant and Ma breather periodically reaches minimum and maximum intensities. & Ma breather is dominant.\\
  \hline
             (ix) & Rogue wave& Akhmediev breather & Rogue wave is dominant with higher amplitude and  Akhmediev breather gets suppressed. & Akhmediev breather is dominant and rogue wave gets suppressed.\\[1ex]
  \hline
\end{tabular}
\end{table}
Apart from this we can have other possible superpositions such as (i)~Two-soliton with Akhmediev breather, (ii)
Two-soliton with Ma breather, (iii)~Two-soliton with Two-soliton, (iv)~Ma breather with Ma breather and (v)~ Akhmediev breather with Akhmediev breather. For brevity we don't discuss these cases here and the interested readers can explore these possibilities in a straightforward way.
\section{Non-autonomous CCGP system}
Next we turn our focus to the following non-autonomous CCGP system:

\bes\bea
i\frac{\partial\psi_1}{\partial t}&=&\left[-\frac{\partial^2}{\partial x^2 }+U_1(x,t)\right]\psi_1-\alpha_1(t)(|\psi_1|^2+2|\psi_2|^2)\psi_1-\alpha_2(t)\psi_2^2\psi_1^*,\\
i\frac{\partial\psi_2}{\partial t}&=&\left[-\frac{\partial^2}{\partial x^2 }+U_2(x,t)\right]\psi_2-\alpha_1(t)(2|\psi_1|^2+|\psi_2|^2)\psi_2- \alpha_2(t)\psi_1^2\psi_2^*,
\eea\label{hccnls}\ees
with time-dependent nonlinearity coefﬁcient $\alpha_j(t)$ and external potential $U(x,t)$.
Here the external potentials $U_1=U_2=U(x,t)$ = $\frac{1}{2}\Omega^2(t)x^2$.
\subsection{Similarity transformation}\label{trans}
First, we look for a similarity transformation that
 transforms the above non-autonomous CCGP system (\ref{hccnls}) into an integrable autonomous CCGP system. This will be of use in identifying the explicit
 forms of the variable nonlinearity coefficient and the nature of
 corresponding trapping potential which can support different nonlinear
 coherent structures in atomic systems.

\indent  For this purpose we introduce the following transformation for the dependent and independent variables in Eqs.~(\ref{hccnls}) with $\alpha_1(t)$=$\alpha_2(t)$=$\alpha(t)$:
\bes\bea
&&\psi_j(x,t) = \xi_1\sqrt{{2\alpha}(t)}~q_j(X(x,t),T(t))~e^{i\varphi(x,t)},\quad j=1,2,
\eea
where
\bea
&&\varphi(x,t) = \left[-\frac{1}{4}\frac{d}{dt}(\ln \alpha)\right]x^2 + \xi_1^2 \xi_2 \left({\alpha}x - \xi_2 \xi_1^2\int {\alpha}^2 dt\right)
\eea
and the new coordinates $X$ and $T$ are defined as
\bea
X = ~\xi_1 \left({\alpha} x - 2\xi_2 \xi_1^2\int {\alpha}^2 dt\right),~~~~~~~~
T = \xi_1^2 \int {\alpha}^2 dt.
\eea\label{str}\ees
Here $\xi_j,~j=1,2,$ are real arbitrary constants.
Then Eq.(\ref{hccnls}) reduces to the following set of
 integrable equations \cite{kannajpa}:
\bes\bea
&iq_{1,T}+ q_{1,XX}+2(|q_1|^2+2|q_2|^2)q_1 + 2q_2^2q_1^*=0,\\
&iq_{2,T}+ q_{2,XX}+2(|q_2|^2+2|q_1|^2)q_2+ 2q_1^2q_2^*=0,
\eea\label{ccnls1}
along with a constraint condition
\bea
\Omega^2(t)=\frac{2\dot{\alpha}^2-\ddot{\alpha}\alpha}{2\alpha^2},
\label{ricatti}
\eea
\ees
where the overdot denotes differentiation with respect to time $t$.
Such type of transformation is possible mainly due to the temporal dependence of the nonlinearity
and that of external harmonic potential; in the absence of such dependence these transformations are not at all possible. Now we can examine the dynamics of different nonlinear waves for a given type of
nonlinearity and potential by expressing Eqs.~(\ref{ccnls1}) as two decoupled NLS equations
using the transformation (\ref{q1}) and then by reconstructing $\psi$ in terms of the original co-ordinates $x$ and $t$ by making use of the similarity transformation (\ref{str}).

\subsection{Forms of time- dependent nonlinear coefficient}
Here we choose two types of nonlinearities namely, periodically
and kink-like modulated nonlinearities, which are of physical interest in BECs.\\

\begin{figure}[H]
\vspace{-0.5cm}
\centering\includegraphics[width=0.5\linewidth]{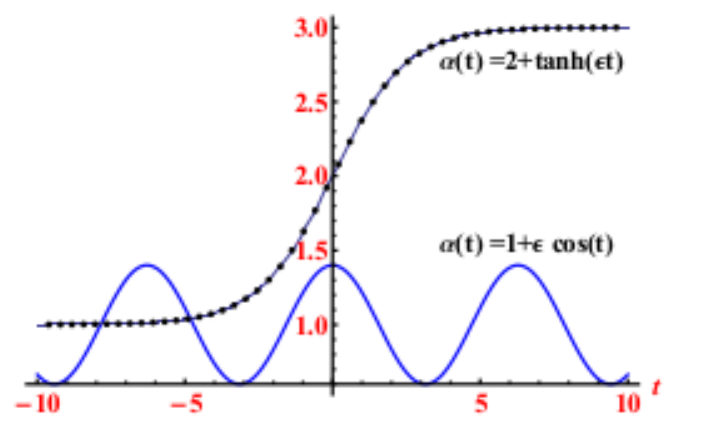}
\caption{Typical form of periodically modulated
and kink-like nonlinearity $\alpha(t)$ for $\epsilon=0.4$.}
\label{fig15}
\end{figure}

\noindent{(a) {\bf Periodically modulated nonlinearity}} \\
 First we choose a periodic form for the nonlinearity coefficient, namely,
 \bes\bea
 \alpha(t) =1+\epsilon\cos(t),
 \label{cos}\eea
 where $\epsilon$ is a real arbitrary parameter.  Such interesting periodic variations
 of the nonlinearity coefficient is considered in BECs \cite{periodic}. This  interesting nonlinearity is plotted in Fig.~(\ref{fig15}). The corresponding expression for $\Omega^2(t)$ is given by:\\
\bea
 \Omega^2(t) =\frac{\epsilon[\cos(t)+\epsilon\cos^2(t)+2\epsilon \sin^2(t)]}{2(1+\epsilon\cos(t))^2}.
 \eea
 \ees

\noindent{(b) {\bf Kink-like nonlinearity}} \\
We also choose another form for variable nonlinearity co-efficient, given by
\bes\bea
 \alpha(t) =2+\tanh(\epsilon t),
 \label{kink}\eea
where $\epsilon$ is an arbitrary real function. In this case, we envision a nonlinearity
that is rapidly varied from one value to another. The evolution of this nonlinearity is shown in Fig.~(\ref{fig15}). Such variation of nonlinearity has been observed in atomic
BECs \cite{pollack}. The corresponding form of $\Omega^2(t)$ is obtained as
\bea
 \Omega^2(t) =\frac{\epsilon^2 sech^2(\epsilon t)[1+2 \tanh(\epsilon t)]}{(2+\tanh(\epsilon t))^2}.
 \eea
 \ees
\subsection{Dynamics of superposed nonlinear waves in non-autonomous CCGP system}
For brevity, here we consider three types of superposed non-autonomous nonlinear waves namely,
superposition of two different non-autonomous solitons, superposition of two-soliton with a non-autonomous rogue wave and superposition of two different non-autonomous rogue waves. The general solution of the non-autonomous CCGP equations (\ref{hccnls}) can be constructed by
using the transformations (\ref{str}) and (\ref{q1}). It can be written as
\bes\bea
\psi_1(x,t) = \xi_1\sqrt{{2\alpha}(t)}~e^{i\varphi(x,t)}(u_1(X,T)+u_2(X,T))/2,\\
\psi_2(x,t) = \xi_1\sqrt{{2\alpha}(t)}~e^{i\varphi(x,t)}(u_1(X,T)-u_2(X,T))/2,
\eea\ees\\
where $u_1(X,T)$ and $u_2(X,T)$ can be any of the five solutions of NLS system (\ref{singlenls}) discussed in Sec. III A, with the replacement of the old-coordinates $x$ and $t$ by the new co-ordinates $X$ and $T$.\\

\noindent{{\bf Case(i)}~ Superposition of two different non-autonomous one solitons}

\begin{figure}[!pht]
\centering\includegraphics[width=0.8\linewidth]{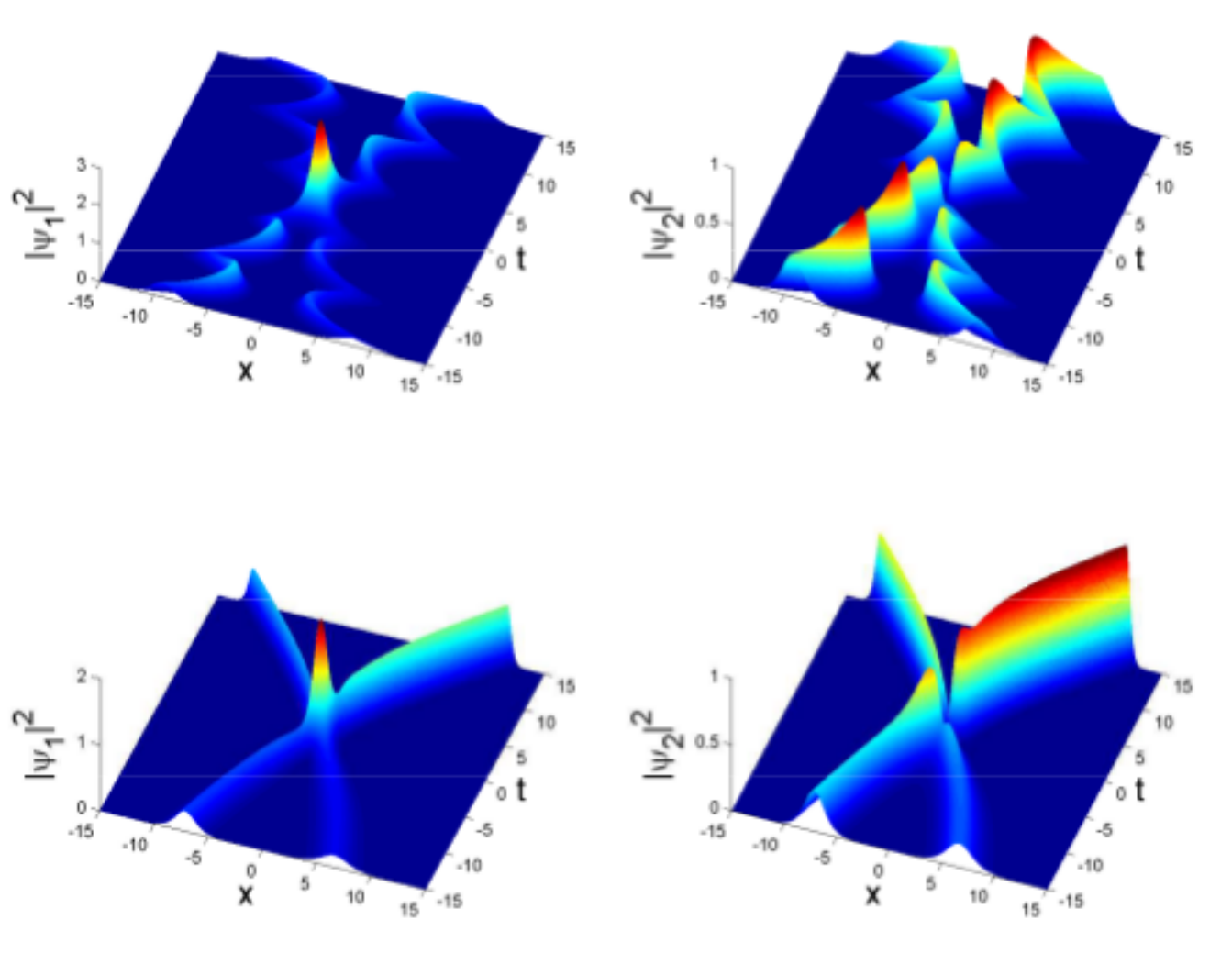}
\caption{Superposition of two different one solitons. Top panel: Periodically modulated nonlinearity for $\epsilon=0.4$, $\xi_1=0.8$ and $\xi_2=0$
; Bottom panel: Kink-like nonlinearity for $\epsilon=0.2$, $\xi_1=0.6$ and $\xi_2=0$.
All other parameters are same as given in the top panel of Fig.~\ref{s-s}.}
\label{fig16}
\end{figure}

Here, we have chosen $u_1$ and $ u_2$ to be of the form given by Eq.~(\ref{soliton}) with $x$ and $t$ replaced by $X$ and $T$, respectively.
The periodically modulated nonlinearity affects both soliton velocity and its shape.
This kind of oscillatory nonlinearity induces oscillations in the soliton profile and results in snake like propagation.
Such non-autonomous solitons can be viewed as ``creeping solitons".
This is shown in the top panel of Fig.~\ref{fig16}.

Next, after introducing the kink nonlinearity, the direction of propagation is also affected (see Eq.~(\ref{kink})), the shape and amplitude of the soliton are altered significantly by the inhomogeneity.
 Another interesting mechanism due to kink-like nonlinearity is ``soliton compression".
 This type of superposition looks similar to the collision of two one solitons.
 Here the amplitudes of the two solitons before collision are lower than that of after collision.
 After collision the width of the solitons (for large positive $t$) get compressed and their amplitudes are enhanced.
Also the separation distance between two solitons before and after
collision is different. In addition to this there is a bending in the path of
the solitons during propagation due to the nature of nonlinearity.
Such a dynamical behaviour is shown in the bottom  panel of Fig.~\ref{fig16}.\\

\noindent{{\bf Case(ii)}~ Superposition of two-soliton with rogue wave}

In this case, we examine the superposition of two-soliton of the form (\ref{twosoliton}) with a rogue
wave (\ref{rogue}) in the presence of periodically modulated nonlinearity. Here, interestingly, the two-solitons are converted into breathers and the rogue wave remains as it is. A distinct feature of this nonlinearity
 is the wings of the breathers are stretched due to oscillation. Note that, here the background is oscillating in
 both the components in a periodic manner as compared with Fig.~\ref{fig11}, where the periodically modulated nonlinearity is absent. This is shown in the top panel of Fig.~\ref{fig17} for illustrative purpose.\\
\begin{figure}[!pht]
\centering\includegraphics[width=0.8\linewidth]{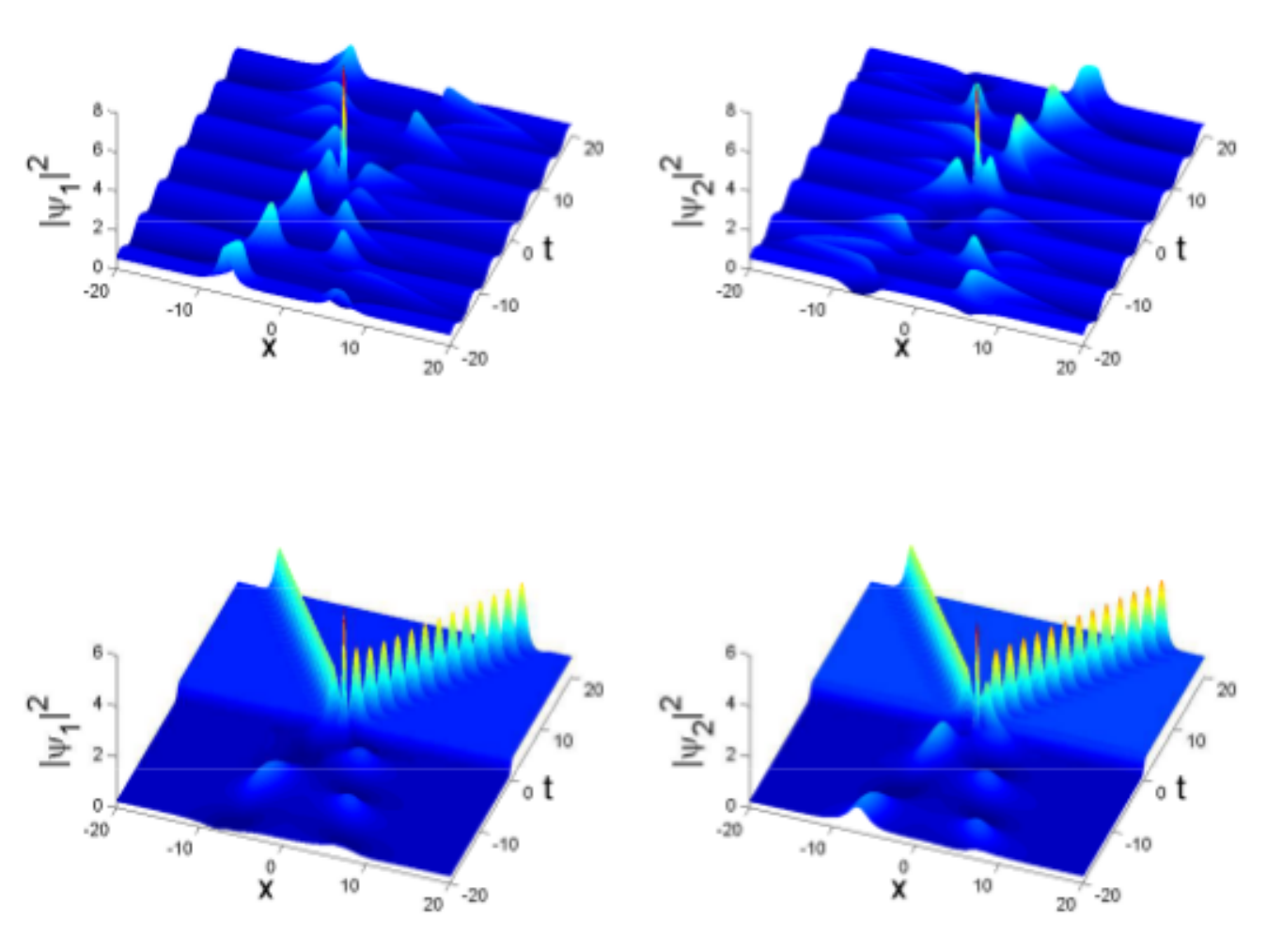}
\caption{Superposition of two-solitons with a rogue wave. Top panel: Periodically modulated nonlinearity
 with parameters $\epsilon=0.5$, $\xi_1=0.8$ and $\xi_2=0$. Bottom panel:
 Kink-like nonlinearity with parameters $\epsilon=2.5$, $\xi_1=0.6$ and $\xi_2=0$.
 All other parameters are same as given in Fig.~\ref{fig11}.}
\label{fig17}
\end{figure}

 Next we consider the superposition of two non-autonomous solitons with a non-autonomous rogue wave in the presence of kink-like nonlinearity (see Eq.~(\ref{kink})). This shows several dynamical features.
 First of all the background does not oscillate instead it displays a step-like profile.
 Here too the two-solitons are converted into breathers. Before collision the
 time period of breathers is greater whereas after collision it decreases and as a result of this
 the period of the breathers is narrowed down. Hence the number of
 breathers has enhanced after collision ($t>0$).
 This is shown in the bottom panel of Fig.~\ref{fig17}.\\

\noindent{{\bf Case(iii)}~ Superposition of non-autonomous rogue waves}\\
By considering two different rogue waves in the presence of periodically modulated nonlinearity as given by Eq. (\ref{cos}), we find that
 the period of oscillation in the background is well diminished as compared to that of Fig. \ref{fig12} (autonomous case). In the second component we observe a non-trivial asymmetric twin peak rogue wave.
\begin{figure}[!pht]
\centering\includegraphics[width=0.8\linewidth]{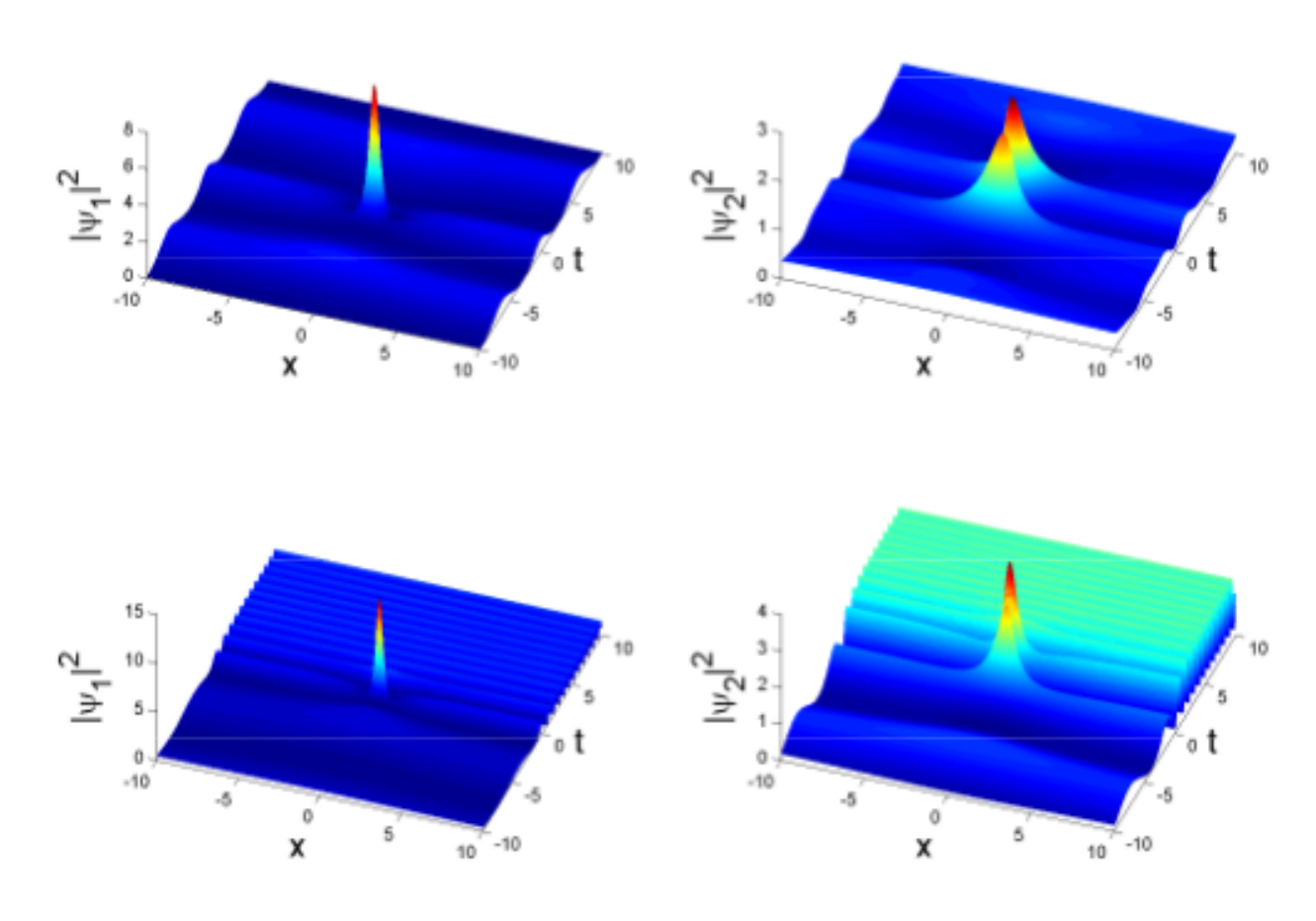}
\caption{Superposition of rogue wave with another rogue wave. Top panel:
Periodically modulated nonlinearity with parameters $\epsilon=0.4$, $\xi_1=0.6$ and $\xi_2=0.1$. Bottom panel:
Kink-like nonlinearity with parameters $\epsilon=0.5$, $\xi_1=0.6$ and $\xi_2=0$.
All other parameters are same as given in Fig.~\ref{fig12}.}
\label{fig17}
\end{figure}

Next we consider the kink-like nonlinearity (see Eq.~(\ref{kink})). Here
the period of oscillating background is greater (almost suppressed) from -$\infty$ to $0$ (see Fig.~\ref{fig12}).
 For $t>0$, the period of oscillation of background is decreased significantly
 with step-like enhancement in its background.
 This is shown in the bottom panel of Fig.~\ref{fig17}. Also the intensity of the rogue wave is increased significantly in the $\psi_1$ and $\psi_2$ components as compared with that of Fig.~\ref{fig12}.

\section{Conclusions}

In conclusion, we have studied the superposed nonlinear waves of coherently coupled GP system. Based on a set of linear transformations,
the coherently coupled GP equations are converted into two decoupled NLS equations. We briefly discuss the available nonlinear wave solutions of scalar NLS equation, namely
solitons (one- and two-), rogue waves and breathers (Ma- and Akhmediev).
We show that the existence of various interesting coherent nonlinear structures such as,
collision of bound soliton with soliton, three soliton collision, coexistence of
rogue wave and Ma breather, coexistence of Ma and Akhmediev breathers, collision of Ma breathers,
bound state of Ma breather by superimposing different types of nonlinear waves. We have also shown that
by such superposition the conventional rogue waves and breathers (Ma/Akhmediev) can be engineered.
Finally, we consider the non-autonomous CCGP system with physically relevant time-dependent nonlinearity coefficients and external potential. With the aid of a similarity transformation, the non-autonomous CCGP system is converted into an integrable autonomous CCGP system with a constraint condition. We examine the two forms
 of nonlinearity coefficient, namely kink-like and periodically modulated function and
 investigate their effects on the novel coherent structures of superposed nonlinear waves. Especially, we demonstrate the possibility of manipulating the nonlinear waves in autonomous as well as non-autonomous settings of the CCGP systems.\\

{\bf Acknowledgments}. The work of T.K. is supported by Department of Science and Technology , Government of India, in the form of a major research project. The authors acknowledge the Principal and Management of Bishop Heber College (Autonomous), Tiruchirappalli for the constant support and encouragement. The authors also thank K. Sakkaravarthi for useful discussions.

\end{document}